\documentstyle[12pt]{article}
\begin{document}
\begin{center}
{\bfseries
FURTHER REMARKS ON ELECTROWEAK MOMENTS OF BARYONS
AND MANIFESTATIONS OF BROKEN SU(3)
\vskip 5mm

S.B. Gerasimov
\vskip 5mm

{\small
{\it
Bogoliubov Laboratory of Theoretical Physics,\\
Joint Institute for Nuclear Research,\\
141980 Dubna (Moscow region), Russia}
}}
\vskip 5mm
\end{center}
\begin{abstract}
The role of nonvalence, {\it e.g.} sea quarks
and/or meson degrees of  freedom in static and quasistatic
baryon electroweak observables,
is discussed within the phenomenological sum rule approach.
The inclusion of nonvalence degrees of freedom in the analysis
of baryon magnetic moments explains extremely strong violation
of the standard $SU(6)$~ symmetry-based quark-model prediction for the
magnetic moment ratio
$R_{\Sigma /\Lambda}=(\Sigma^{+} + 2\Sigma^{-})/(-\Lambda)
\simeq .23$, while the value $R_{\Sigma /\Lambda}(SU(6))= 1$
corresponds to the nonrelativistic quark model.
We also obtain $F/D=.72$ for the quark-current-baryon coupling
$SU(3)_{f}$ ratio. The implications for the "strangeness" magnetism of the
nucleon and for weak axial-to-vector coupling constant
relations measured in the lowest octet baryon $\beta$-decays are
discussed. The latter shows up the possible role
of the induced second-class form factor (the "weak-electricity", or
pseudotensor form factor) in the extraction of the $(g_1/f_1)$-values
from the hyperon's~ $\beta$~- decay data.
\end{abstract}
\vskip 8mm

{\bf 1.}In this report we present some further
consequences from sum rules for the static
electroweak characteristics of baryons following mainly from the
phenomenology of broken internal symmetries.
The phenomenological sum rule techniques was chosen to
obtain a more reliable, though not very much detailed information about
the hadron properties in question.
The main focus was laid on the role of nonvalence degrees of
freedom ( the nucleon sea partons and/or peripheral meson currents )
in parameterization and description of hadron magnetic moments and
axial-vector coupling constants.

As is known, in the broken $SU(3)$-symmetry approach, based on the
non-relativistic quark model (NRQM) of the ground state octet baryons \cite{Ge66},
where $B\leftrightarrow 2q_{even}+q_{odd}$, and
the magnetic moments of constituent quarks
in the corresponding baryons $B=\{P,N;\Sigma^{\pm};\Xi^{o,-};\Lambda\}$,
satisfy the relation $\mu(u):\mu(d):\mu(s)=-2:1:(m_{d}/m_{s})$,
one obtains the familiar expressions
for magnetic moments
\begin{eqnarray}
\mu(B)\equiv B &=&(4/3)q_{e} - (1/3)q_{o},\nonumber\\
\Lambda &=& s,\nonumber \\
\mu(\Lambda \Sigma^{o}) &=& (1/\sqrt{3})(u-d)
\label{NRQM}
\end{eqnarray}
(herewith,
we use the particle and quark symbols the for corresponding magnetic moments).
The most spectacular difficulty of the above parameterization
is seen from comparing two ratios
$R_{\Sigma/\Lambda}$\cite{Lip99} and $R_{\Xi/\Lambda}$~with experimental
values~\cite{PDG}--the first one is drastically broken, while both should be valid
in the NRQM:
\begin{eqnarray}
R_{\Sigma/\Lambda}=\frac{\Sigma^{+}+2\Sigma^{-}}{-\Lambda}=\frac{s(\Sigma)}{s(\Lambda)}=
1~{\it vs}~.23~\cite{PDG},\nonumber \\
R_{\Xi/\Lambda}=\frac{\Xi^{o}+2\Xi^{-}}{4\Lambda}=\frac{s(\Xi)}{s(\Lambda)}=
1~{\it vs}~1.04\cite{PDG}.
\label{Ratios}
\end{eqnarray}
Earlier we considered a number of
consequences of sum rules for the static
electroweak characteristics of baryons following from the theory of broken
internal symmetries and common features of the quark models
including  corrections due to nonvalence degrees of freedom -- the sea
partons and/or the meson clouds at the periphery of baryons
and no assumptions referring to the nonrelativistic quark dynamics were made.

Here, we list some of the earlier discussed~
\cite{Ge88,Ge95,Ge96,Ge97} sum rules
(we use the particle and quark symbols for the corresponding magnetic moments):
\begin{eqnarray}
\alpha_D = {D\over F+D}|_{mag} = {1\over 2}(1-{\Xi^0 - \Xi^- \over \Sigma^+ - \Sigma^-
- \Xi^0 + \Xi^-}).
\label{F-D}
\end{eqnarray}
The $D$- and $F$- constants in Eq.(\ref{F-D}) parameterize the "reduced" matrix
elements of the quark current operators where $SU(3)$ symmetry-breaking
effects are contained in the factorized effective coupling constants
of the single-quark-type operators, while other contributions
({\it e.g.} representing the pion exchange current effects) are
cancelled in all sum rules by construction.
The ratio $u/d \ne -2$
\begin{eqnarray}
{u \over d} = {\Sigma^+ (\Sigma^+ - \Sigma^-) - \Xi^0 (\Xi^0 - \Xi^-) \over
\Sigma^- (\Sigma^+ - \Sigma^-) - \Xi^- (\Xi^0 - \Xi^-)},
\label{u-d}
\end{eqnarray}
is related to the chiral constituent quark
model where a given baryon consists of three "dressed" massive
constituent quarks. Owing to the virtual transitions
$q \leftrightarrow  q + \pi(\eta), q \leftrightarrow K + s$,
the "magnetic anomaly" is developing, {\it i.e.,}
$u/d = -1.80 \pm .02 \neq Q_{u}/Q_{d}=-2$.

The ratio $s/d \simeq .64$ demonstrating the $SU(3)$-symmetry breaking
is evaluated via
\begin{eqnarray}
{s \over d} = {\Sigma^+ \Xi^- - \Sigma^- \Xi^0 \over \Sigma^- (\Sigma^+ - \Sigma^-)
- \Xi^- (\Xi^0 - \Xi^-)}
\label{s-d}
\end{eqnarray}
Now, we list some consequences of the obtained sum rules.
The numerical relevance of the adopted parameterization is seen from
the results
enabling even estimation from one of the obtained sum rules, namely,
\begin{eqnarray}
&&(\Sigma^+ - \Sigma^-) (\Sigma^+ + \Sigma^- - 6\Lambda + 2\Xi^0 + 2\Xi^-) \nonumber \\
&&- (\Xi^0 - \Xi^-) (\Sigma^+ + \Sigma^- + 6\Lambda - 4\Xi^0 - 4\Xi^-) = 0,
\label{L-sr}
\end{eqnarray}
the necessary effect of the isospin-violating $\Sigma^{o} \Lambda$-mixing.
By definition, the $\Lambda$--value
entering into Eq.(\ref{L-sr}) should be "refined" from the
electromagnetic $\Lambda\Sigma^0$--mixing affecting  $\mu(\Lambda)_{exp}$.
Hence, the numerical value of $\Lambda$, extracted from Eq.(\ref{L-sr}),
can be used to determine the $\Lambda\Sigma^o$--mixing angle through the
relation
\begin{eqnarray}
\sin \theta_{\Lambda\Sigma} \simeq \theta_{\Lambda\Sigma} =
{\Lambda -\Lambda_{exp} \over 2\mu(\Lambda\Sigma)} = (1.43 \pm 0.31) 10^{-2}
\label{mix}
\end{eqnarray}
in accord with the independent estimate of
$\theta_{\Lambda\Sigma}$ from
the electromagnetic mass-splitting sum rule \cite{DvH}.

Naturally, our approach is free of the disbalance problem exemplified
in Eqs.(\ref {Ratios}).
With the parameters
$u/d=-1.80$ and $\alpha_{D}=(D/(F+D))_{mag}=0.58$, defined without
including the $\Lambda$-hyperon magnetic moment in fit
and taking into account the $\Sigma^{o}-\Lambda$~-mixing, we obtain
$R_{\Sigma / \Lambda} \simeq .27$~ and $R_{\Xi / \Lambda} \simeq 1.13$,
which turn out to be in excellent accord with data
if one takes also into account in Eq.(\ref{Ratios}) the $\Lambda$~-value
corrected for mixing:~~$\Lambda_{0}\simeq -.567~ n.m.$.
For further use, we also list below the limiting relations following
from the neglect of the nonvalence degrees of freedom
\begin{eqnarray}
\Sigma^+ [\Sigma^-] = P[-P-N] +
(\Lambda - {N\over 2})( 1+{2N \over P}),\nonumber \\
\Xi^0[\Xi^-] = N[-P-N] + 2(\Lambda -
{N\over 2})( 1+ {N\over 2P}),\nonumber \\
\mu(\Lambda\Sigma) = -{\sqrt{3}\over 2}N.
\label{qNR}
\end{eqnarray}
We stress that no
NR assumption or explicit $SU(6)$-wave function are used this time.
The ratio $F/D = .64$ in this case and it is definitely less than
$ F/D=.72$,~ when nonvalence degrees of freedom are included.
This is the demonstration of substantial influence of the nonvalence
degrees of freedom on this important parameter.

{\bf 2.}One can note that the accordance of the ratios $R_{\Sigma/\Lambda,
\Xi/\Lambda}$ with
data is valid in two, seemingly dual, parameterizations of
the baryon magnetic moments. The first is specified by the renormalization
of the constituent quark characteristics by the meson current effects
resulting in $u/d \ne -2$,~etc. However,
one can follow a complementary view of the nucleon structure,
keeping the constraint u/d=-2,~and the OZI-rule violating the contribution
of sea quarks parameterized as~
$\Delta (N)= \sum_{q=u,d,s} \mu(q)<N|\bar{s}s|N> \ne 0 $.

We have referred to this approach \cite{Ge95} as a correlated current-quark picture
of nucleons and made use of it to estimate the contributions of the sea
quarks to baryon magnetic moments. In particular,
the following important sum rules were obtained (all quantities are in n.m.):
\begin{eqnarray}
 \Delta(N) = {1\over 6}(3(P+N) - \Sigma^+  + \Sigma^- -\Xi^0 +\Xi^-)&=&
-.06\pm.01,\nonumber \\
 \mu_N(\overline ss)=\mu(s)
\langle N|\overline ss|N\rangle=
(1- {d\over s})^{-1}\Delta(N) &=& .11 \pm .02,\\
G^{s}_{M}(0)= -\frac{1}{2}(1- {d\over s})^{-1}(3(P+N) - \Sigma^+  + \Sigma^- -\Xi^0 +\Xi^-)
&=&~-.33 \pm .06
\label{str-spin}
\end{eqnarray}
where the ratio d/s=1.55 follows
from the correspondingly modified Eq.(\ref{s-d}) (that is with
$Y$ replaced by $(Y - \Delta(N))$).
By definition, $\mu_N(\overline ss)$ represents the contribution of
strange ("current") quarks to nucleon magnetic moments.
Actually, our Eq. (\ref{str-spin}) is equivalent to the half-sum
of two relations in Ref.\cite{Le98}
where the ratios of effective magnetic moments of quarks in different baryons
should be taken the same.
Indeed, within the lattice QCD approach with a chosen extrapolation
prescription to the chiral limit of small current quark masses~\cite{Le98}
two sum rules were written down and the numerical estimation obtained
\begin{eqnarray}
G^{s}_{M}(0)&=&-(1-\frac{d}{s})^{-1}[2P+N-\frac{u(P)}{u(\Sigma^{+})}
(\Sigma^{+}-\Sigma^{-})]\nonumber\\
G^{s}_{M}(0)&=&-(1-\frac{d}{s})^{-1}[P+2N-\frac{u(N)}{u(\Xi^{o})}
(\Xi^{o}-\Xi^{-})]\\
G^{s}_{M}(0)&=&-.16 \pm .18
\end{eqnarray}
At last, as the representative of the approach pretending to be the limit
of the QCD with a large number of colours $N_{C}\to \infty$, we
write also the sum rule of the chiral soliton model~\cite{Hong01}
\begin{eqnarray}
G^{s}_{M}(0)= \frac{1}{3}(N-\Sigma^{+}-4\Sigma^{-}+\Xi^{o}-3\Xi^{-})=+.32
\end{eqnarray}
Within still rather large experimental uncertainties, the latest value
of the SAMPLE Collaboration~\cite{SAMPLE-00}
\begin{equation}
G^{s}_{M}(0)|_{exp}=.01\pm .29\pm .31\pm.07,
\end{equation}
where the three errors are statistical, systematic and theoretical,~
respectively,~
does not contradict any of the model values mentioned above.

It is quite natural to expect that we have now the evident
constraint $G^{s}_{M}(0)\to 0$ in the limit when we neglect all
nonvalence quark contributions to baryon magnetic moments,
that is when all the relations of Eq.(\ref{qNR}) are put into
any of the sum rules for $\mu_{N}(s\bar{s})$.
We notice that our relation for
$\mu_{N}(s\bar s)$ and $G^{s}_{M}(0)$ satisfies this constraint identically,~
and the lattice QCD relations~\cite{Le98} require the "environment"
influence to be absent, {\it i.e.} $(u(P)/u(\Sigma^{+})=(u(N)/u(\Xi^{o})
=1$, while the chiral soliton relation~\cite{Hong01} requires the fulfillment
of the substantially stronger additional assumption $\Lambda =-(N/2)$
which is equivalent to exact $SU(3)$-symmetry relations for magnetic
moments. This peculiarity makes the last relation less attractive
and, theoretically, more subject to doubts compared to the first two
 predicting the negative value of $G^{s}_{M}(0)$.

{\bf 3.}To estimate a possible influence of the SU(3) breaking in the ratio of the
weak axial-to-vector coupling constants, we adopt the following
prescription suggested by the success of our parameterization of the
baryon magnetic moment values within the constituent quark model.
In essence,
we assume that the leading symmetry breaking effect is
produced by different renormalization
of the $\bar q q W$- strangeness-conserving and strangeness-nonconserving
vertices with the participation of the constituent quarks.
We note further that in all but one \cite{Hs88} analyses of
the hyperon $\beta$-decays,  the absence of the "weak electricity"
form factor $g_2(Q^2)$ due to the induced second-class weak current
has been postulated from the very beginning.
However, the fit to all
$\Sigma^{-} \rightarrow ne\bar {\nu}$ decay data of Ref. \cite{Hs88}
 with $g_2 \neq 0$ yields $g_a=(g_1/f_1)
 -.20 \pm .08$ and $(g_2/f_1)=+.56 \pm .37$.
It seems that one cannot then define $(F/D)_{\Delta S=1}$ because data
for all other  hyperons have been treated under the assumption $g_2=0$.

Having in mind the evidence of a potentially important correlation between
the values of the axial-to-vector coupling
$(g_{1}/f_{1})$ and
the "weak-electricity"-to-vector $(g_{2}/f_{1})$ coupling ratio, observed in the
$\Sigma^{-}\to Ne\nu$~-decay~\cite{Hs88}, we parameterize
$(g_{i}/f_{1}), i=1,2$ in the strangeness-violating $\beta$-decays
by their (different) $F_{i}$-~and $D_{i}$ parameters in the expression
\begin{equation}
\frac{g_{1}}{f_{1}}(F_{1},D_{1})+r_2\frac{g_{2}}{f_{1}}(F_{2},D_{2})
=\frac{g_{1}}{f_{1}}(F_{1}^{eff},D_{1}^{eff})
\end{equation}
with the same correlation coefficient $r_2 \simeq -.25$, quoted in
the recent review\cite{CSW03} for both $\Sigma^{-}$~and $\Lambda$ semileptonic
decays but not measured in the $\Xi^{o,-}$~-decays yet.
The $F_{1}^{eff}$~and~$D_{1}^{eff}$~ will then play the role of
the "effective" parameters defined from data with the {\it ad hoc} constraint
$g_{2}=0$. Taking $F_{1}+D_{1}=1.26$, and $F_{1}/D_{1}|_{\Delta S=0}=.72$
we find $F_{2}$~ and $D_{2}$~ from the known data on
the $\Sigma^{-}\to N$~ and
$\Lambda \to P$~semileptonic decays
\begin{eqnarray}
F_{1}-D_{1}+r_2(F_{2}-D_{2})=-.34\pm.02,\\
F_{1}+(1/3)D_{1}+r_2[F_{2}+(1/3)D_{2}]=.718\pm .015,
\end{eqnarray}
to obtain "effective" parameters for the $\Xi^{-}$~ and $\Xi^{o}$ decays
equal to~ $.19\pm.03~(.25\pm.05)$ and~ $1.25\pm .03~(1.32\pm .20)$, respectively.
The presently measured "effective" parameters~\cite{CSW03}
are given in the parentheses
and they are seen to be within one standard deviation
from the calculated ones. We also notice that
the ratio of~ $|(g_{2}/f_{1}|$ in the $\Sigma^{-}$-~and $\Lambda$-decays
is close to that calculated within the dynamical model of Ref.~\cite{PD73};~
however, the same type ratios including the $\Xi^{-,o}$-decay constants are
completely different. Accumulation of new data announced in \cite{CSW03}
and their improved analysis is, therefore, of great interest.

{\bf 4.} To conclude, besides the importance of resolution of
the problem on the presence
and quantitative role of the weak second-class current and the
corresponding form factors in the hyperon $\beta$-decay
observable, one can also mention major theoretical
interest in the careful study of the strangeness-conserving
$\Sigma ^{\pm} \rightarrow \Lambda e^{\pm} \nu(\bar{\nu})$
transitions which would not only prove (or disprove) hypotheses
about the dependence of $(F/D)$-ratios on $\Delta S$, labelling
the transitions, but also would provide information on the isospin
breaking effects underlying the $\Lambda - \Sigma^{o}$ - mixing.

\end{document}